\documentstyle[prl,aps,twocolumn,amssymb,graphicx]{revtex}
\def\bea{\begin{eqnarray}}
\def\eea{\end{eqnarray}}

\renewcommand\epsilon{\varepsilon}
\def\beq{\begin{equation}}
\def\eeq{\end{equation}}

\def\lsim{\mathrel{\raise.3ex\hbox{$<$\kern-.75em\lower1ex\hbox{$\sim$}}} }
\def\gsim{\mathrel{\raise.3ex\hbox{$>$\kern-.75em\lower1ex\hbox{$\sim$}}} }

\begin{document}
\draft
\title{Leptogenesis-MNS Link in Unified Models with Natural Neutrino Mass
Hierarchy}
\author{S.F.King$^{(a)}$ \thanks{E-mail:sfk@hep.phys.soton.ac.uk}\\
\small   $^{(a)}$Department of Physics and Astronomy,
University of Southampton, Southampton, SO17 1BJ, U.K.}
\maketitle
\thispagestyle{empty}
\begin{abstract}
We discuss the relation between leptogenesis and the MNS phases in the class
of see-saw models called light sequential dominance in which the
right-handed neutrinos dominate sequentially, with the dominant
right-handed neutrino being the lightest one. 
The heaviest right-handed neutrino then decouples,
leaving effectively two right-handed neutrinos.
Light sequential dominance is motivated by
$SO(10)$ models which predict three right-handed neutrinos.
With an approximate Yukawa texture zero 
in the 11 position there are only two see-saw phases 
which are simply related to the MNS phase 
measurable at a neutrino factory and the
leptogenesis phase. The leptogenesis phase is predicted to be
equal to the neutrinoless double beta decay phase.
\end{abstract}
 \narrowtext
There is by now good evidence for neutrino masses and mixings
from atmospheric and solar neutrino oscillations \cite{Gonzalez-Garcia:2002sm}.
The minimal neutrino sector required to account for the
atmospheric and solar neutrino oscillation data consists of
three light physical neutrinos with left-handed flavour eigenstates,
$\nu_e$, $\nu_\mu$, and $\nu_\tau$, defined to be those states
that share the same electroweak doublet as the left-handed
charged lepton mass eigenstates.
The neutrino flavor eigenstates $\nu_e$, $\nu_\mu$, and $\nu_\tau$ are
related to the neutrino mass eigenstates $\nu_1$, $\nu_2$, and $\nu_3$
with mass $m_1$, $m_2$, and $m_3$, respectively, by a $3\times3$ 
unitary matrix called the MNS matrix $U_{MNS}$
\cite{Maki:1962mu,Lee:1977qz}
\begin{equation}
\left(\begin{array}{c} \nu_e \\ \nu_\mu \\ \nu_\tau \end{array} \\ \right)=
U_{MNS}
\left(\begin{array}{c} \nu_1 \\ \nu_2 \\ \nu_3 \end{array} \\ \right)
\; .
\end{equation}
Assuming the light neutrinos are Majorana,
$U_{MNS}$ can be parameterized in terms of three mixing angles
$\theta_{ij}$, a Dirac phase $\delta_{\rm MNS}$, together with
two Majorana phases $\beta_1,\beta_2$, as follows
\begin{equation}
U_{MNS}=R_{23}U_{13}R_{12}P_{12}
\end{equation}
where
\bea
R_{23} & = &
\left(\begin{array}{ccc}
1 & 0 & 0 \\
0 & c_{23} & s_{23} \\
0 & -s_{23} & c_{23} \\
\end{array}\right), \ \  
R_{12}  = 
\left(\begin{array}{ccc}
c_{12} & s_{12} & 0 \\
-s_{12} & c_{12} & 0\\
0 & 0 & 1 \\
\end{array}\right) \\
U_{13} & = &
\left(\begin{array}{ccc}
c_{13} & 0 & s_{13}e^{-i\delta_{\rm MNS}} \\
0 & 1 & 0 \\
-s_{13}e^{i\delta_{\rm MNS}} & 0 & c_{13} \\
\end{array}\right), \\
P_{12}  & = &
\left(\begin{array}{ccc}
e^{i\beta_1} & 0 & 0 \\
0 & e^{i\beta_2} & 0\\
0 & 0 & 1 \\
\end{array}\right)
\label{MNS}
\eea
where $c_{ij} = \cos\theta_{ij}$ and $s_{ij} = \sin\theta_{ij}$.

The most elegant explanation of small neutrino masses continues
to be the see-saw mechanism \cite{seesaw,Mohapatra:1979ia}. 
According to the see-saw
mechanism, lepton number is broken at high energies due to 
right-handed neutrino Majorana masses, resulting in small left-handed
neutrino Majorana masses suppressed by the heavy mass scale.
The see-saw mechanism also provides an attractive mechanism
for generating 
the baryon asymmetry of the universe 
via leptogenesis \cite{yanagida1}, \cite{luty}.

There is a large literature concerned with a possible 
link between the phases responsible for leptogenesis and those
in the low energy neutrino sector, especially the Dirac phase 
$\delta_{\rm MNS}$ which may be measured at a neutrino factory.
In general the conclusion seems to be that there is no direct link 
\cite{large}.
The leptogenesis phase appears to be quite independent of the 
neutrino factory phase. However in specific regions of parameter
space \cite{Davidson:2002em} or in specific models 
\cite{Joshipura:2001ui,2RHN,Frampton:2002qc} such a link may exist.
For example in models with only two right-handed neutrinos, 
there does appear to be a link at least in special cases 
\cite{2RHN,Frampton:2002qc}.
However many models predict three right-handed neutrinos due to
a gauged $SU(2)_R$ which may be embedded into a larger gauge
group such as $SO(10)$. Nevertheless it is well known 
\cite{King:1998jw,King:1999mb,King:2002nf} that certain classes of 
three right-handed neutrino models can behave effectively as two right-handed
neutrino models. This provides the motivation for our present study.

The purpose of the present paper is to discuss leptogenesis in
a class of models with three right-handed neutrinos in which 
a neutrino mass hierarchy arises naturally due to a
single right-handed neutrino dominantly contributing to the
heaviest neutrino mass, and a 
second right-handed neutrino dominantly contributing to the
the second heaviest neutrino mass \cite{King:1999mb}.
This mechanism called sequential dominance was discussed recently 
including phases in \cite{King:2002nf} from which many of the 
results in this paper have been derived. We shall further
focus on the case that the dominant right-handed neutrino is the
lightest one, called light sequential dominance (LSD).
The physical motivation for LSD is that in such models the neutrino Yukawa
matrix has the same universal form as the quark Yukawa matrices,
with no large off-diagonal elements, and such theories 
are compatible with $SO(10)$ \cite{King:2001uz}.
The main physical consequence of LSD
is that the heaviest right-handed
neutrino is irrelevant for both neutrino oscillation experiments
and leptogenesis, and effectively decouples, leading to an 
effective two right-handed neutrino description.

For the present discussion we shall work in the flavour basis
where the charged lepton mass matrix and the right-handed neutrino
Majorana mass matrix are both diagonal with real positive eigenvalues.
We write the latter as
\begin{equation}
M_{RR}=
\left( \begin{array}{ccc}
Y & 0 & 0    \\
0 & X & 0 \\
0 & 0 & X'
\end{array}
\right) 
\label{seq1}
\end{equation}
where we assume 
\begin{equation}
Y\ll X\ll X'.
\label{LSD}
\end{equation}
We write the neutrino Yukawa matrix as
\begin{equation}
Y^{\nu}_{LR}=
\left( \begin{array}{ccc}
0 & a & a'    \\
e & b & b' \\
f & c & c'
\end{array}
\right),
\label{dirac}
\end{equation}
in the convention where the Yukawa matrix
in Eq.\ref{dirac} corresponds to the Lagrangian coupling
$\bar{L}HY^{\nu}_{LR}\nu_R$, where $L$ are the left-handed lepton
doublets, $H$ is the Higgs doublet, and $\nu_R$ 
are the right-handed neutrinos. 
The condition for sequential dominance (SD) is \cite{King:1999mb}
\beq
\frac{|e^2|,|f^2|,|ef|}{Y}\gg
\frac{|xy|}{X} \gg
\frac{|x'y'|}{X'}
\label{srhnd}
\eeq
where $x,y\in a,b,c$ and $x',y'\in a',b',c'$, 
where all Yukawa couplings are assumed to be complex.
The combination of Eqs.\ref{LSD},\ref{srhnd} is called LSD.
As we shall see LSD implies that the third right-handed neutrino
of mass $X'$ is irrelevant for both leptogenesis and neutrino 
oscillations. In addition many realistic $SO(10)$ models involve an
approximate texture zero in the 11 position \cite{King:2001uz}, 
and we assume that.
This will have the effect of removing one
of the see-saw phases.

Assuming Eq.\ref{srhnd} the neutrino masses
are given to leading order in $m_2/m_3$ by \cite{King:2002nf},
\bea
m_1 & \sim & O(\frac{x'y'}{X'}v_2^2) \label{m1} \\
m_2 & \approx &  \frac{|a|^2}{Xs_{12}^2} v_2^2 \label{m2} \\
m_3 & \approx & \frac{|e|^2+|f|^2}{Y}v_2^2 
\label{m3}
\eea
where $v_2$ is a Higgs vacuum expectation value (vev) associated with
the (second) Higgs doublet that couples to the neutrinos
and $s_{12}=\sin \theta_{12}$ given below. 
Note that with SD each neutrino mass is generated
by a separate right-handed neutrino, and the sequential dominance condition
naturally results in a neutrino mass hierarchy $m_1\ll m_2\ll m_3$.
The neutrino mixing angles are given to leading order in $m_2/m_3$ by
\cite{King:2002nf},
\bea
\tan \theta_{23} & \approx & \frac{|e|}{|f|} \label{23}\\
\tan \theta_{12} & \approx &
\frac{|a|}
{c_{23}|b|
\cos(\tilde{\phi}_b)-
s_{23}|c|
\cos(\tilde{\phi}_c)} \label{12} \\
\theta_{13} & \approx &
e^{i(\tilde{\phi}+\phi_a-\phi_e)}
\frac{|a|(e^*b+f^*c)}{[|e|^2+|f|^2]^{3/2}}
\frac{Y}{X}
\label{13}
\eea
where we have written some (but not all) complex Yukawa couplings as
$x=|x|e^{i\phi_x}$. The phase $\delta_{\rm MNS}$
is fixed to give a real angle
$\theta_{12}$ by,
\beq
c_{23}|b|
\sin(\tilde{\phi}_b)
\approx
s_{23}|c|
\sin(\tilde{\phi}_c)
\label{chi1}
\eeq
where 
\bea
\tilde{\phi}_b &\equiv & 
\phi_b-\phi_a-\tilde{\phi}+\delta_{\rm MNS}, \nonumber \\ 
\tilde{\phi}_c &\equiv & 
\phi_c-\phi_a+\phi_e-\phi_f-\tilde{\phi}+\delta_{\rm MNS}
\label{bpcp}
\eea
The phase $\tilde{\phi}$
is fixed to give a real angle
$\theta_{13}$ by \cite{King:2002nf},
\beq
\tilde{\phi} \approx  \phi_e-\phi_a -\phi_{\rm COSMO}
\label{phi2dsmall}
\eeq
where
\beq
\phi_{\rm COSMO}=\arg(e^*b+f^*c). 
\label{lepto0}
\eeq
is the leptogenesis phase 
corresponding to the interference diagram involving the
lightest and next-to-lightest right-handed neutrinos
\cite{Hirsch:2001dg,King:2002nf}.
Note that the sign of baryon asymmetry
for the LSD case of interest here 
is then determined by $Y_B\propto +\sin 2\phi_{COSMO}$ \cite{Hirsch:2001dg}.
However it should be pointed out that in the LSD case here
with Eq.\ref{LSD} successful thermal
leptogenesis requires $Y\approx 10^{12}{\rm GeV}$
\cite{Buchmuller:2002jk} which would lead to a serious
problem with the production of TeV mass gravitinos in supersymmetric models.
Returning to Eq.\ref{lepto0}, it may be expressed as
\beq
\tan \phi_{\rm COSMO} \approx 
\frac{|b|s_{23}s_2+|c|c_{23}s_3}{|b|s_{23}c_2+|c|c_{23}c_3}.
\label{phi121}
\eeq
Inserting $\tilde{\phi}$ in Eq.\ref{phi2dsmall}
into Eqs.\ref{chi1},\ref{bpcp},
\bea
&&c_{23}|b|\sin(\eta_2+\phi_{\rm COSMO}+\delta_{\rm MNS}) \nonumber \\
& \approx &
s_{23}|c|
\sin(\eta_3+\phi_{\rm COSMO} +\delta_{\rm MNS}).
\label{chi12}
\eea
Eq.\ref{chi12} may be expressed as
\beq
\tan (\phi_{\rm COSMO}+\delta_{\rm MNS}) \approx 
\frac{|b|c_{23}s_2-|c|s_{23}s_3}{-|b|c_{23}c_2+|c|s_{23}c_3}
\label{phi12del}
\eeq
where we have written $s_i=\sin \eta_i, c_i=\cos \eta_i$
where
\beq
\eta_2\equiv \phi_b-\phi_e, \ \ \eta_3\equiv \phi_c-\phi_f
\label{eta}
\eeq
are invariant under a charged lepton phase transformation.
The reason that the see-saw parameters only
involve two invariant phases $\eta_2, \eta_3$ rather than the usual six
is due to the LSD 
assumption which has the effect of decoupling the  
heaviest right-handed neutrino, which removes three phases, 
together with the assumption
of a 11 texture zero, which removes another phase.

Eq.\ref{phi12del} shows that
$\delta_{\rm MNS}$ is a function of 
the two see-saw phases
$\eta_2 , \eta_3$ that also determine $\phi_{\rm COSMO}$ in Eq.\ref{phi121}.
If both the phases $\eta_2 , \eta_3$ are zero,
then both $\phi_{\rm COSMO}$ and $\delta_{\rm MNS}$ are necessarily zero.
This feature is absolutely crucial. It means that,
barring cancellations, measurement of a non-zero value for 
the phase $\delta_{\rm MNS}$ at a neutrino factory will be a signal of a
non-zero value of the leptogenesis phase $\phi_{\rm COSMO}$.

So far we have not discussed the Majorana phases $\beta_1$, $\beta_2$ 
which appear in Eq.\ref{MNS}. In LSD the lightest neutrino mass $m_1$
is very small which effectively makes $\beta_1$ unmeasurable.
The remaining Majorana phase is given from the relation 
\beq
\delta_{\rm MNS}+\beta_2= \frac{\phi_2}{2}-\frac{\phi_3}{2}
\label{beta2}
\eeq
where $\phi_2$, $\phi_3$ are the phases of the neutrino masses 
$m_2'=m_2e^{i\phi_2}$, $m_2'=m_2e^{i\phi_3}$, after the neutrino mass
matrix has been diagonalised. 
In our conventions these phases 
are removed to leave the positive neutrino masses in Eq.\ref{m3}
resulting in Eq.\ref{beta2}. We find
\beq
\phi_2=2\phi_a, \ \ \phi_3=2(\phi_a+\phi_{\rm COSMO}).
\label{phi}
\eeq
From Eqs.\ref{beta2},\ref{phi}, we find 
\beq
\phi_{\rm COSMO}=-(\delta_{\rm MNS}+\beta_2),
\label{remarkable1}
\eeq
The combination of phases
in Eq.\ref{beta2} is just the phase $\phi_{\beta \beta 0\nu}$ 
which enters $|m_{ee}|$ in neutrinoless double beta decay, 
when expressed in terms of oscillation parameters.
The magnitude of this phase is
\beq
|\phi_{\beta \beta 0\nu}|=|\delta_{\rm MNS}+\beta_2|.
\label{beta}
\eeq
From Eq.\ref{remarkable1} and \ref{beta} we find the remarkable result
\beq
|\phi_{\rm COSMO}|=|\phi_{\beta \beta 0\nu}|.
\label{remarkable2}
\eeq
In the bottom-up analysis of \cite{Davidson:2002em}, 
they also found a relation like Eq.\ref{remarkable2} in a particular region of 
their low energy parameter space.

If we were to assume {\it ad hoc} that a second Yukawa element were to be 
set to zero, then another invariant see-saw phase could be removed, and then 
both $\phi_{\rm COSMO}$ and $\delta_{\rm MNS}$ would be determined
in terms of a single see-saw phase. In this case there would be 
a direct relation between $\phi_{\rm COSMO}$ and $\delta_{\rm MNS}$.
For example if we set $c=0$ then the only remaining see-saw phase is 
$\eta_2$ and Eqs.\ref{phi121},\ref{phi12del} would imply that 
$\phi_{\rm COSMO}=\eta_2=\phi_b-\phi_e$ and 
\beq
\delta_{\rm MNS}=-2\phi_{\rm COSMO}.
\label{signs}
\eeq
The sign of the CP asymmetry phase $\delta_{\rm MNS}$ measurable
at a neutrino factory is then predicted to be negative since the baryon number
of the universe $Y_B\propto +\sin 2\phi_{COSMO}$ is positive. 
The LSD assumption in our model Eqs.\ref{LSD},\ref{srhnd}
means that the heaviest right-handed
neutrino of mass $X'$ is irrelevant for both leptogenesis and
for determining the neutrino masses and mixings, so the model
reduces effectively to one involving only two right-handed neutrinos,
as first observed in \cite{King:1999mb}. In fact,
assuming $c=0$, our model in Eqs.\ref{seq1},\ref{LSD},\ref{dirac},\ref{srhnd} 
reduces effectively to the case of two right-handed neutrinos with
$Y\ll X$ considered in the phenomenological analysis in \cite{Raidal:2002xf}.
A similar two right-handed neutrino model with
$X\ll Y$ proposed in \cite{Frampton:2002qc} corresponds to the limiting 
three right-handed neutrino case $X\ll Y \ll X'$
which is not as well motivated as LSD. 
Another possiblity is to have heavy sequential
dominance (HSD) corresponding to $X'\ll X \ll Y$, leading to 
the ``lop-sided'' neutrino Yukawa matrix. In the HSD case
the lightest right-handed neutrino of mass $X'$
is responsible for leptogenesis but has couplings and
phases which are completely irrelevant to the neutrino mass and
mixing spectrum. Therefore in HSD there is no
leptogenesis-MNS link \cite{King:2002nf}, although $X'$ may be
light enough to be consistent 
with the production of TeV mass gravitinos in supersymmetric models
\cite{Hirsch:2001dg}.

To conclude, we have discussed the relation between leptogenesis 
and the MNS phases in the LSD class of models defined by
Eqs.\ref{seq1},\ref{LSD},\ref{dirac},\ref{srhnd}.
Although the class of model looks quite specialised,
it is in fact extremely well motivated since it 
allows the neutrino Yukawa matrix to have the same universal form
as the quark Yukawa matrices 
consistent with $SO(10)$ for example \cite{King:2001uz}.
The large neutrino mixing angles and neutrino mass hierarchy
then originate naturally
from the sequential dominance mechanism
without any fine tuning \cite{King:1999mb}.
Within this class of models we have shown that
the two see-saw phases $\eta_2$, $\eta_3$ 
are related to $\delta_{\rm MNS}$ and $\phi_{\rm COSMO}$
according to Eqs.\ref{phi121},\ref{phi12del}.
Remarkably, the leptogenesis phase is predicted to be
equal to the neutrinoless double beta decay phase as in
Eq.\ref{remarkable2}. Since the heaviest right-handed
neutrino of mass $X'$ is irrelevant for both leptogenesis and
for determining the neutrino masses and mixings, the model
reduces effectively to one involving only two right-handed neutrinos
\cite{King:1999mb}. With an additional {\it ad hoc} assumption
that $c=0$ the model then further reduces to a case considered
in \cite{Raidal:2002xf}, and we find
a direct relation between $\phi_{\rm COSMO}$ and $\delta_{\rm MNS}$
as in Eq.\ref{signs}. In this case the fact that the universe is made 
of matter not antimatter then predicts the sign of $\delta_{\rm MNS}$
measurable at a neutrino factory to be negative.

I thank Alejandro Ibarra for enlightening discussions and for
critically reading the manuscript. I also thank John Ellis, Graham Ross
and Liliana Velasco-Sevilla for discussions.
I would also like to thank the CERN Theory Division for its hospitality
and PPARC for a Senior Research Fellowship.

%

\end{document}